# SENSOR TECHNOLOGIES IN CANCER RESEARCH FOR NEW DIRECTIONS IN DIAGNOSIS AND TREATMENT: AN EXPLORATORY ANALYSIS

**Mario Coccia**

CNR -- National Research Council of Italy,
Collegio Carlo Alberto, Via Real Collegio, 30-10024 Moncalieri (Torino), Italy
E-mail: mario.coccia@cnr.it

**Saeed Roshani**

Allameh Tabataba'i University, Faculty of Management and Accounting,
Department of Technology and Entrepreneurship Management, Tehran, Iran
E-mail: Roshani@atu.ac.ir

**Melika Mosleh**

Birmingham Business School, College of Social Sciences, University of Birmingham, UK
E-mail: mxm1219@alumni.bham.ac.uk

**Abstract:**

The goal of this study is an exploratory analysis concerning main sensor technologies applied in cancer research to detect new directions in diagnosis and treatments. The study focused on types of cancer having a high incidence and mortality worldwide: breast, lung, colorectal and prostate. Data of the Web of Science (WOS) core collection database are used to retrieve articles related to sensor technologies and cancer research over 1991-2021 period. We utilized Gephi software version 0.9.2 to visualize the co-word networks of the interaction between sensor technologies and cancers under study. Results show main clusters of interaction per typology of cancer. Biosensor is the only type of sensor that plays an essential role in all types of cancer: breast cancer, lung cancer, prostate cancer, and colorectal cancer. Electrochemical sensor is applied in all types of cancer under study except lung cancer. Electrochemical biosensor is used in breast cancer, lung cancer, and prostate cancer research but not colorectal cancer. Optical sensor can also be considered one of the sensor technologies that significantly is used in breast cancer, prostate cancer, and colorectal cancer. This study shows that this type of sensor is applied in more diversified approaches. Moreover, the oxygen sensor is mostly studied in lung cancer and breast cancer due to the usage in breath analysis for the treatment process. Finally, Cmos sensor is a technology used mainly in lung cancer and colorectal cancer. Results here suggest new directions for the evolution of science and technology of sensors in cancer research to support innovation and research policy directed to new technological trajectories having a potential of accelerated growth and positive social impact for diagnosis and treatments of cancer.

## 1. Introduction

The research field of sensor is undergoing a significant change to support the evolution of science and technologies in society (Andersen et al., 2004; Coccia et al., 2021; Coccia and Watts, 2020; Coccia, 2019, 2019a, 2021, 2021a; Rao et al., 2018; Wilson, 2004). The goal of this study is an exploratory analysis to detect main sensor technologies applied in cancer research for improving diagnosis and treatments and reducing whenever possible mortality between countries. The vast literature in these topics shows main results for cancer research (Bayford et al., 2022; Kaur et al., 2022; Li et al., 2020, 2020a; Rey-Barth et al., 2022; Sivanandhan et al., 2022; Thakare et al., 2022). As far as breast cancer is concerned, Wu et al. (2022) design a dual-aptamers functionalized gold for classification of breast cancer based on Förster resonance energy transfer, which is potentially useful for quantitative classification of different subtypes of breast cancer. Lu et al. (2022) argue that phthalates can penetrate the environment and enrich various aquatic organisms through the food chain, which is involved in promoting the growth of breast cancer. It is of current interest to develop new sensors for phthalates. Results show that guest-induced reassembly brings forth significant fluorescence change, which is a promising way of designing new fluorescent probes for the analysis of phthalates in the environment and food. Pothipor et al. (2022) show that a dual-mode electrochemical biosensor is successfully developed for simultaneous detection of two different kinds of breast cancer biomarkers. The experimental results suggest that this label-free biosensor exhibits good linear responses to the concentrations of both target analytes with the limits of detection. This assay strategy has a great potential to be further developed for the simultaneous detection of a variety of miRNAs and protein biomarkers for point-of-care diagnostic applications. Kim et al. (2022) maintain that mechanophores are molecular motifs that respond to mechanical perturbance with targeted chemical reactions toward desirable changes in material properties. Taking advantage of the strengths of mechanophores and high-intensity focused ultrasound, mechanochemical dynamic therapy can provide noninvasive treatments for diverse cancer types (cf., Mohan et al., 2022). About prostate cancer, Bax et al. (2022) argue that diagnostic protocol is affected by poor accuracy and high false-positive rate and propose an electronic nose for non-invasive prostate cancer detection. The approach proved to be effective in mitigating drift on 1-year-old sensors by restoring accuracy from 55% to 80%, achieved by new sensors not subjected to drift. The model achieved, on double-blind validation, a balanced accuracy of 76.2%. Prema et al. (2022) examine the biological synthesis of gold nanoparticles using green tea and their cytotoxicity against human prostate cancer cells. The findings suggest that the biosynthesized reduced prostate cancer cell proliferation and exert their anti-proliferative action on the prostate cancer cell line by inhibiting growth, decreasing DNA synthesis, and triggering apoptosis. In lung cancer, Joshi et al. (2022) argue that proper and early diagnosis of cancers are a basic to efficient treatment and better prognosis and report a simple and label-free method of detection of two antigens: carcinoembryonic antigen (CEA) and cytokeratin-19 fragment (CYFRA 21-1) that are the biomarkers of many cancers including lung cancer. The responses of the sensors ranged from 10.96 to 26.48% for 0.25 pg/mL to 20 ng/mL CEA and it varied from 17.66 to 26.68% for 0.25 pg/mL to 20 ng/mL CYFRA 21-1. Kaya et al. (2022) review the recent advances and improvements (2011–2021 period) in nanomaterials based electrochemical biosensors for the detection of the lung and colon cancer biomarkers (cf., Tumuluru et al., 2022). In colon cancer, Jiang et al. (2022) argue that Transient receptor potential vanilloid 1 (TRPV1) acts as cellular sensor and is implicated in the tumor microenvironment cross talk and the functional role of TRPV1 in colorectal cancer (CRC). The study reveals an important role for TRPV1 in regulating the immune microenvironment during colorectal tumorigenesis and might be a potential target for CRC immunotherapy. Welz et al. (2022) point out that the intestinal epithelium undergoes constant self-renewal from intestinal stem cells. Together with genotoxic stressors and failing DNA repair, this self-renewal causes susceptibility toward malignant transformation. The study shows that X-box binding protein 1 is a stress sensor involved in coordinating epithelial DNA damage responses and stem cell function.

In this context of the evolution of sensor technologies towards manifold fields of research, the motivation of this study is to clarify the role of research and technology in sensors for main typologies of cancer, describing the networks of interconnection of sensors with other technologies and scientific aspects related to cancer under study. Proposed methodology can indicate new directions of sensor technologies for diagnosis and treatments of cancer and help policymakers to allocate with efficiency financial resources to support scientific and technological development in these critical fields of research in society (cf., also Ardito et al., 2021; Coccia, 2018, 2019a, 2020; Coccia and Finardi, 2012, 2013; Kashani and Roshani, 2019; Roshani et al., 2021).

The balance of the paper proceeds as follows. First, it describes the data and methodology, applying a novel information processing approach of computational scientometrics, to generate maps of science that can explain the ecosystem and evolution of sensor research and technologies in specific typologies of cancer. We then show the results and conclude with a discussion on new directions of the evolution of sensor technology in cancer and limitations of the study to be solved with future studies.

## 2. Materials and Methods

First, this study focuses on main typologies of cancer that have the highest estimated age-standardized incidence and mortality rate worldwide as indicated in Table 1 and Figure 1 based on data by Globocan (2020).

Table 1. Estimated age-standardized incidence and mortality rates (World) in 2020, worldwide, both sexes, all ages

| Cancer | Incidence | Mortality |
|---|---|---|
| Breast | 47.8 | 13.6 |
| Prostate | 30.7 | 7.7 |
| Lung | 22.4 | 18 |
| Colorectum | 19.5 | 9 |
| Cervix uteri | 13.3 | 7.3 |
| Stomach | 11.1 | 7.7 |
| Liver | 9.5 | 8.7 |
| Corpus uteri | 8.7 | 1.8 |
| Ovary | 6.6 | 4.2 |
| Thyroid | 6.6 | 0.43 |

Data source: Globocan (2020)

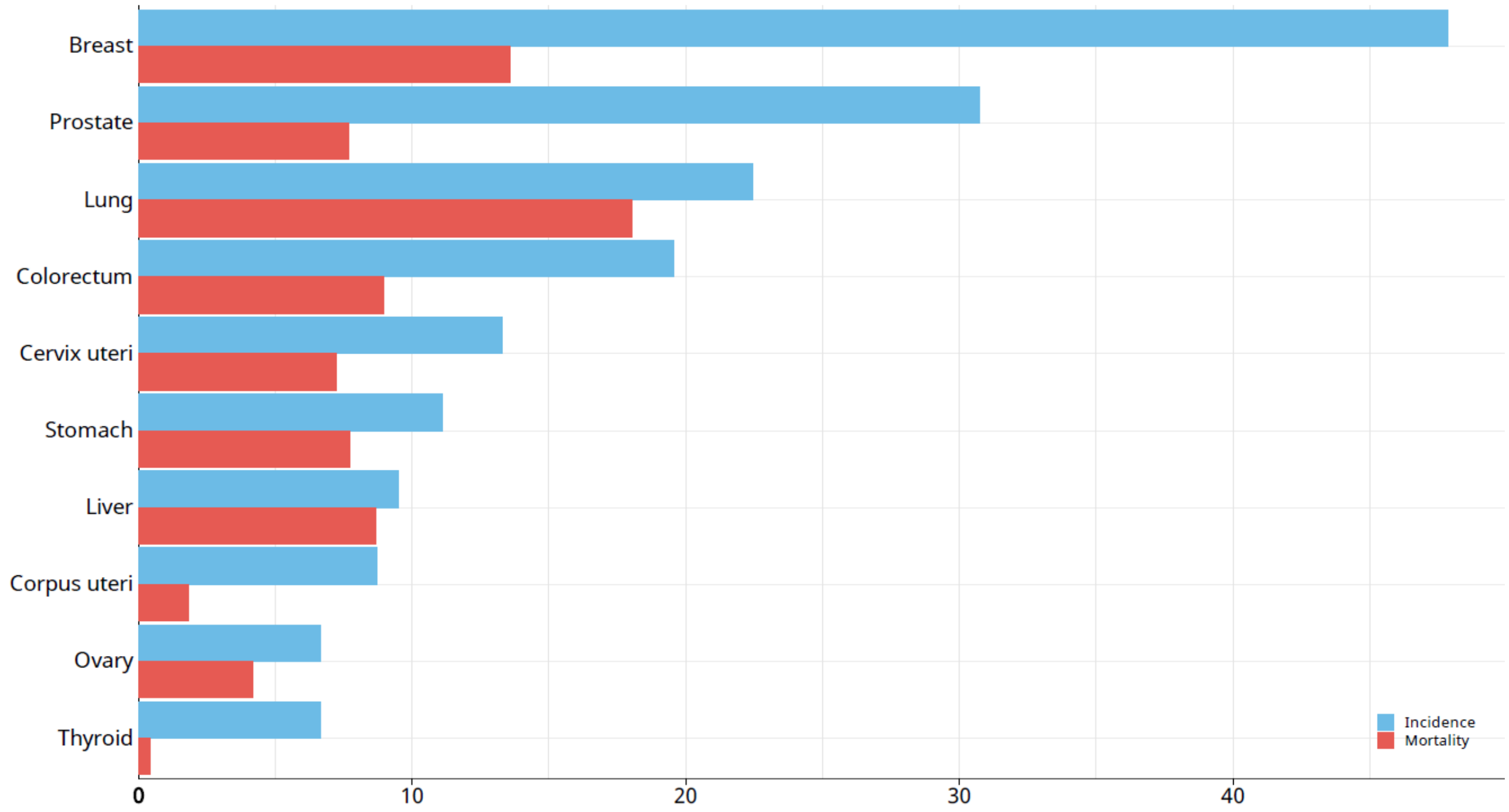

Figure 1. Estimated age-standardized incidence and mortality rates (World) in 2020, worldwide, both sexes, all ages. Red bars indicate mortality; blue ones incidence. Data source: Globocan (2020)

Considering the data just mentioned, this study focuses on the investigation of sensor technology and research in the following four main types of cancer:

- Breast cancer
- Lung cancer
- Prostate cancer
- Colorectal cancer

### 2.1. Data sources and retrieval strategy

To address the main question of this study, we used the Web of Science (WOS, 2021) core collection database to retrieve the articles related to sensor technologies and cancers. In this study, we focused as said on four major types of cancers for a comparative analysis (Coccia, 2018a) including: Breast cancer (BC), lung cancer (LC), Prostate cancer (PC) and Colorectal cancer (CC). We used the following strategy for extracting related articles for further processing. The term "sensor" was searched with "breast cancer" in the topics of articles. We changed the "breast" to "lung" for extracting the articles related to lung cancer, "prostate" for prostate cancer and "colorectal" for colorectal cancer. The results are refined by document type = "Articles", Language = "English", Publication years = (1991-2021), Web of Science index = "SCI-EXPANDED"). We found 1117 unique articles for breast cancer, 764 articles for lung cancer, 454 articles for prostate cancer, and 282 articles for colorectal cancer.

### 2.2. Data processing procedure

To find the applications of sensors technology in cancers considered in this article, we used the original keywords (DEs) provided by the authors as the basis for constructing the word co-occurrences networks. According to this technique, two terms are considered co-occurrence whenever they simultaneously appear in a single document (Delecroix and Epstein, 2004). To find the relevant sensor technologies to each of the cancer we studied, we applied the following procedure:

Keywords standardization: We tried to clean the keywords according to their meaning and structures in this step. For instance, we combined the "bio-sensor" and "biosensor" into Biosensor. Also, we changed all abbreviations and plural forms of nouns into a basic form (e.g., computers changed to computer and "AMPK" changed to "Amp-activated Protein kinase".

Network construction: We used SCI2 tool V. 1.3 for constructing the word co-occurrences network (Sci2 Team, 2009). As mentioned above, we used the article's original keywords (DE tag) to create the networks.

We create four different breast, lung, Prostate, and colorectal cancer networks. Also, we removed the isolated nodes after analyzing the nodes and links in the networks.
Path-finding: path-finding is a technique for choosing the shortest links between two nodes. To reduce the links of our networks and emphasize the most important nodes, we used a minimum spanning tree (MST) path-finder algorithm (Quirin et al., 2008). All the calculations are implemented by SCI2 tool. Also, as an input parameter of the algorithm, we set the parameter Weight Attribute measures to "SIMILARITY" and Edge Weight Attribute to "Unweighted". The initial network of breast cancer contains 10,149 links, and links reduced network contains 2,318 edges. The lung cancer network contains 6,128 initially and has 1539 links after applying the link reduction algorithm. After the algorithm implementation, the Prostate cancer network had 3,395 links and held 850 edges. These results for colorectal cancer include 2,394 links at the first and 528 links after implementing the link reduction algorithm.

Visualizations: We utilized Gephi software (Bastian et al., 2009) version 0.9.2 to visualize the networks The nodes indicate the original keywords, and links show the co-occurrences among them. Also, the size of nodes is based on the Betweenness centrality. This measure is used for detecting the nodes which have a connecting role in the network (Kashani and Roshani, 2019). We set the Betweenness centrality as an indicator for identifying the groups of the path in each network. In other words, nodes with Betweenness centrality greater than 0.1 are considered thresholds for identifying the groups.

## 3. Results and Discussions

### 3.1. Breast cancer

Breast cancer has seriously threatened women health in the world (Chagpar and Coccia, 2019; Coccia, 2013, 2019c,1029b, 2020; Harbeck and Gnant, 2017). As mentioned earlier, the breast cancer sample contains 1,117 articles, and it is the most extensive dataset in our analysis. This network contains 2,319 nodes (keywords) and 2,318 links. Also, this network includes 149 sensors that are interconnected to the other nodes. Figure 2 shows the breast cancer co-word network.

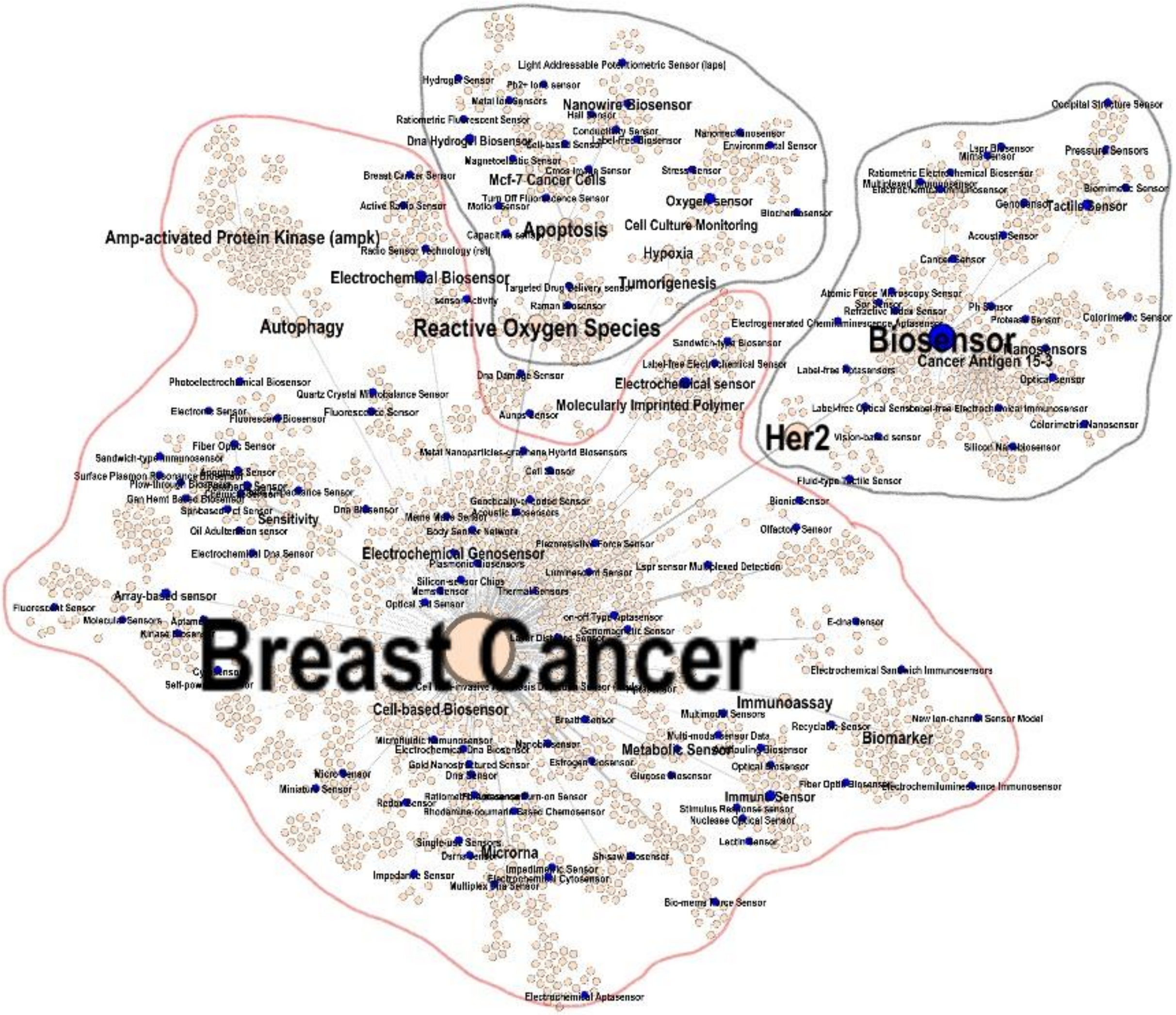

Figure 2. Co-word analysis map of breast cancer.

Based on the Betweenness centrality value, we found three sub-groups in this network. Table 1A in Appendix shows the most important information about these groups. Also, there are 149 sensors interconnected to this network.

In group 1 related to breast cancer, we have five hot topics: Autophagy, Immunoassay, Electrochemical Biosensor, Tumorigenesis, and Microrna, which have a high co-occurrence with Electrochemical Biosensor, Electrochemical Sensor, Oxygen Sensor, Immuno sensor, and Array-based Sensor.

Group 2 with head of Biosensor has a couple of hot topics, including Her2, Cancer antigen, Nanoparticle, Tactile Sensor, and Signal Amplification, which are significantly related to Optical sensor, Colorimetric Sensor, Colorimetric Nanosensor, Ph. Sensor, and Refractive Index Sensor.

Group 3 led by Reactive Oxygen Species, it is related to Apoptosis, Mcf-7 Cancer Cells, Circulating Tumor Cell, Dna Hydrogel Biosensor, and Carbon Dot. In this domain, Raman Biosensor, DNA Hydrogel Biosensor, Light Addressable Potentiometric Sensor (laps), Capacitive Sensor, and Label-free Biosensor are couple of sensors which is included in this group of connection (see in Appendix Table 1A about groups, core keywords and related sensors in the breast cancer network).

### 3.2. Lung cancer

According to the size of the network, the lung cancer is the second cancer were analyzed (cf., Coccia, 2012; 2014; 2017; 2019d). This network contains 1,540 nodes (keywords) and 1,539 links. Also, this network includes 121 sensors that interconnected to the other nodes. Figure 3 shows the lung cancer co-word network.

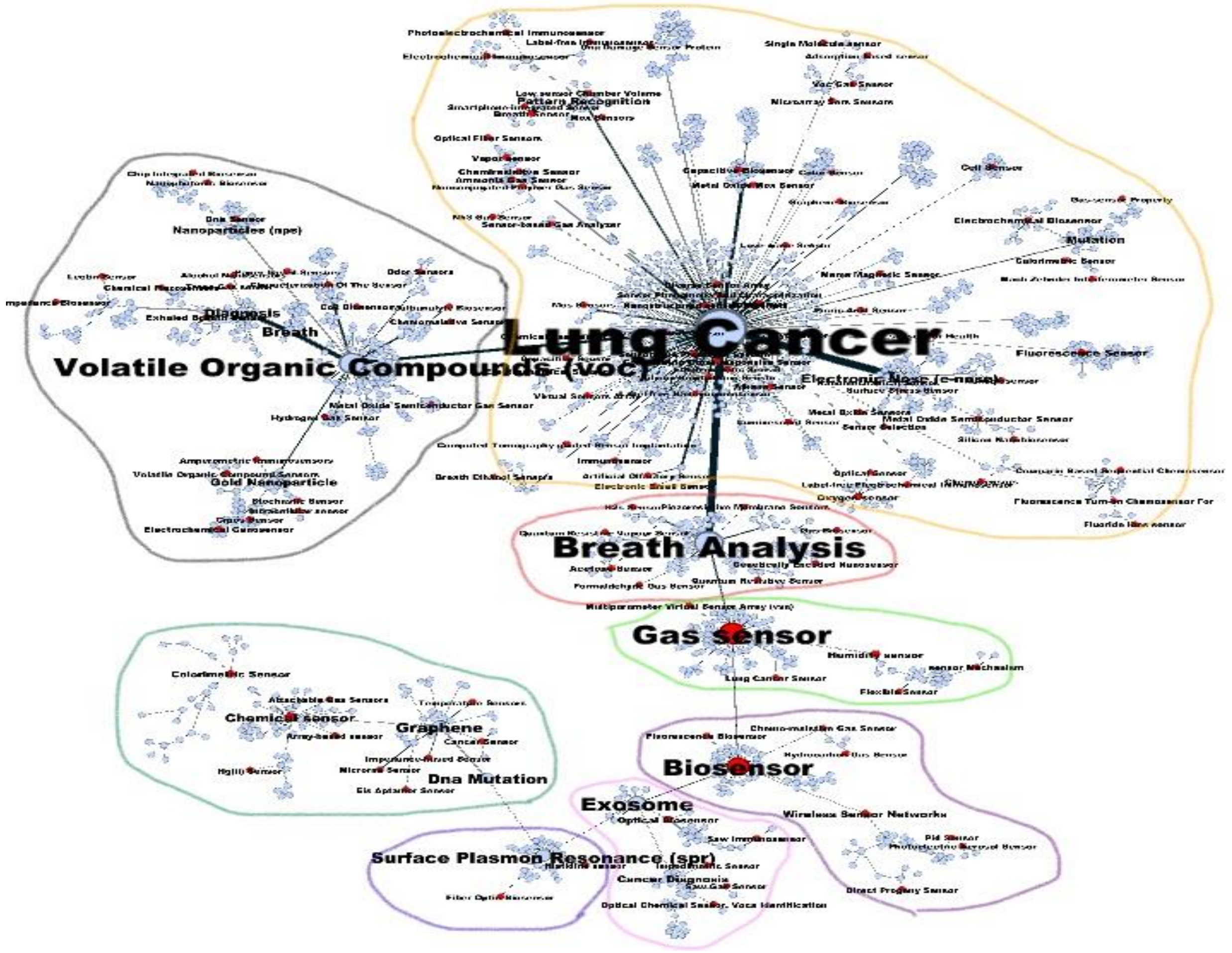

Figure 3. Co-word analysis map of lung cancer

As shown in figure 3, Regarding the Betweenness centrality threshold of 0.1, this network has eight path groups. All nodes with a value of Betweenness centrality greater than 0.1 are considered the head of a group. Top keywords and all the sensors included in these categories have been shown in Table 2A in Appendix.

Group 1 with the head of lung cancer is related to 5 hot topics: Electronic Nose (e-nose), Pattern Recognition, Mutation, Nrf2 (Nuclear factor-erythroid factor 2-related factor 2), and Calix[4]arene. It is also illustrated that 69 sensors are interconnected to this path. The oxygen sensor, Electrochemical Biosensor, Cell Sensor, Metal Oxide Semiconductor Sensor, Chemiresisitve Sensor, Breath Sensor, Electrochemical Immunosensor, and Capacitive Biosensor are a couple of sensors that are involved in the creation of linkage path among these nodes.

In group 2 led by Volatile Organic Compounds (voc), we identified 23 related sensors including, Exhaled Breath Sensor, Dna Sensor, Volatile Organic Compound Sensors, Cmos Sensor, Electrochemical Genosensor, etc. The hot topics related to this group of nodes are Breath, Diagnosis, Gold Nanoparticle, Nanoparticles, and Exhaled Breath.

Group 3 connected to Breath analysis, has 8 interconnected sensors, including Piezoresistive Membrane Sensors, H2s Sensor, Formaldehyde Gas Sensor, Quantum Resistive Vapor Sens Quantum Resistive Sensor, Acetone Sensor, and Genetically Encoded Nanosensor. The most frequent keywords involved in this path group are: Real-Time, Diabetes, Non-invasive, Formaldehyde, and Sno2.

Group 4 with the head of Gas sensor is related to important keywords of Humidity sensor, Health Monitoring, Reduced Graphene Oxide, and Acetone. Six sensors of gas, lung cancer, Sensor, multiparameter virtual sensor array, humidity sensor, sensor Mechanism, and Flexible Sensor are part of this path group.

In Group 5, Biosensor with the most important technology in creating the path among different sensors and keywords, it is responsible for making a bridge between the other seven sensors of Fluorescence Biosensor, Wireless Sensor Networks, Photoelectric Aerosol Sensor, Direct Progeny Sensor, Pid Sensor, Chemo-resistive Gas Sensor and most frequent keywords of Wireless Sensor Networks, Radon, Smart Home, Electrochemical Inhibitors, and Hydrocarbon Gas Sensor.

The next group, led by Exosome connected to the top five keywords of Cancer Diagnosis, Optical Biosensor, Immunoassay, and Multi-wall Carbon Nanotubes related to five sensors of Optical Biosensor, Saw Gas Sensor, Optical Chemical Sensor, Vocs Identification, Saw Immunosensor, and Impedimetric Sensor.

The 7$^{th}$ group distinguished by Surface Plasmon Resonance (spr) includes five hot keywords of Endoscopy, Au Nps, Signal Enhancement, Erlotinib, and Tollen's Reagent, which have two sensors of  Histidine sensor and Fiber Optic Biosensor in their category.

In group 8 related to Graphene, we have five DNA Mutation, Chemical Sensor, Urine Headspace, Colorimetric Sensor, and Tuberculosis keywords. Ten sensors of Chemical Sensor, Colorimetric Sensor, Cancer Sensor, Array-based Sensor, Temperature Sensors, Impedance-based Sensor, Attachable Gas Sensors, Hg(ii) Sensor, Eis Aptamer Sensor, and Microrna Sensor are the important sensing technologies in this group.

### 3.3. Prostate cancer

This network contains 870 nodes (keywords) and 869 links. This network includes 72 sensors that are interconnected to the other nodes. Figure 4 shows the prostate cancer co-word network.

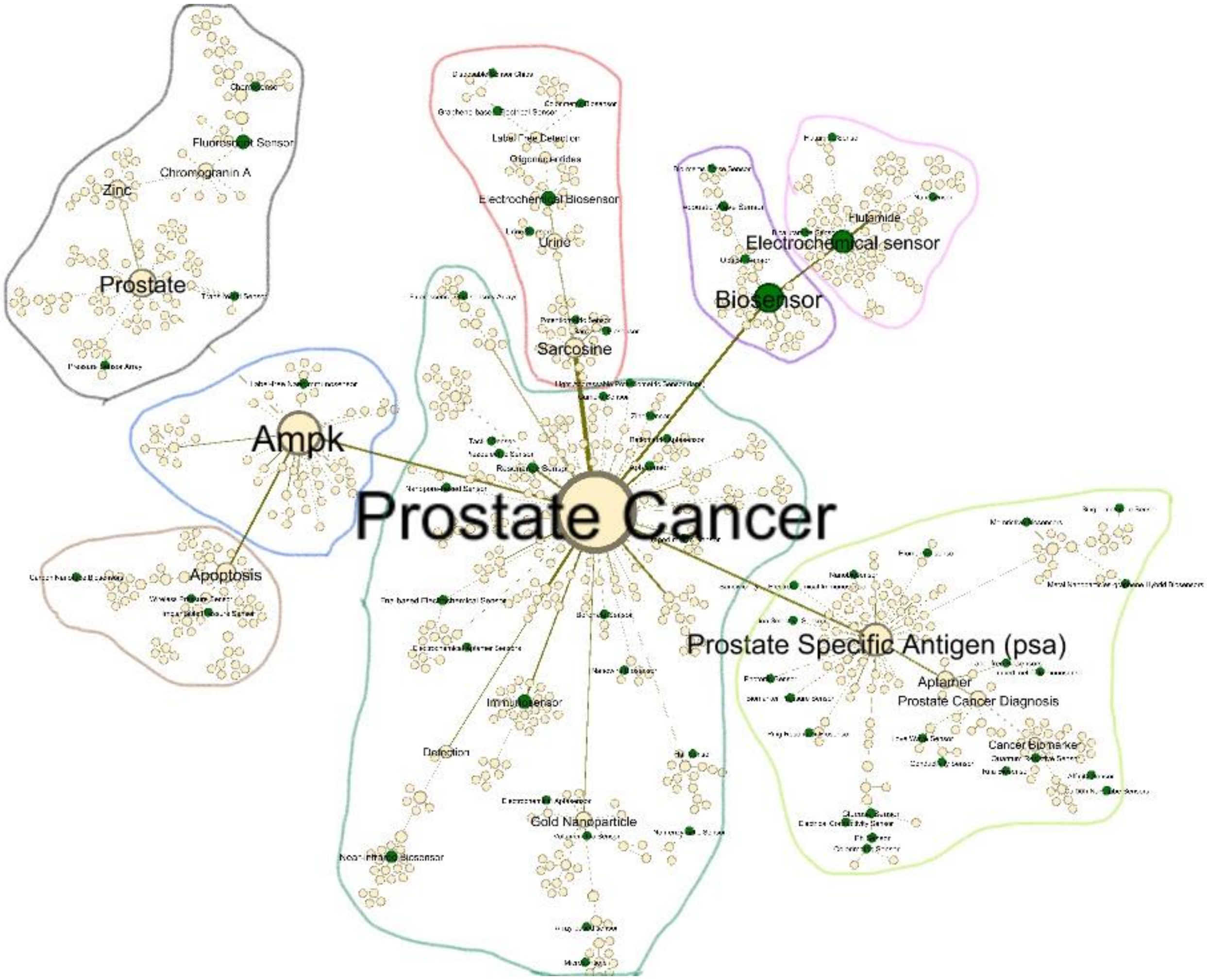

Figure 4. Co-word analysis map of Prostate cancer

Figure 4 shows that there are eight groups in the Prostate cancer network. The most important nodes based on the Betweenness centrality value are: Prostate cancer, Prostate-specific antigen, Activated Protein Kinase (AMPK) and Biosensor. Table 3A in Appendix shows the groups, core keywords of them and related sensors.

Group 1, led by Prostate cancer, is related to 5 hot topics of Gold Nanoparticle, Immunosensor, Detection, Porcine Liver Esterase, and Fluorescent Probe. It is also illustrated that 27 sensors are interconnected to this path. Immunosensor, Near-infrared Biosensor, Resonance Sensor, Fna-based Electrochemical Sensor, Impedimetric Sensor, and Aptasensor are a couple of sensors involved in the creation of linkage path among these nodes.

In group 2, led by Activated Protein Kinase (Ampk), we identified Label-free Nanoimmunosensor as the only connected Sensor. The hot topics related to this group of nodes are Apoptosis, Lkb1, Lung Cancer, Mtorc1, Castrate-resistant

Group 3, connected to Prostate Specific Antigen (PSA), has 22 interconnected sensors, including Glucose Sensor, Ph Sensor, Colorimetric Sensor, Quantum Resistive Sensor, etc. The most frequent keywords involved in this path group are Aptamer, Prostate Cancer Diagnosis, Cancer Biomarker, and Silicon Nanowire.

Group 4 with the head of Biosensor is related to important keywords of Antibody, Prostatic Carcinoma, DNA, Acoustic Wave Sensor, Cancer Metastasis. Four sensors of Biosensor, Optical Sensor, Acoustic Wave Sensor, Bio-mems Force Sensor are part of this group.

The next group, led by Prostate connected to the top five Zinc, Chromogranin A, Fluorescent Sensor, Imaging Diagnosis, and Peptide related to four sensors of Trans-rectal and Pressure Sensor Array, Chemosensor, and Fluorescent Sensor.

In Group 6, Electrochemical Sensor with the most important position in creating the path among different sensors and keywords is responsible for making a bridge between Nanosensor, Flutamide Sensor, Bicalutamide Sensor and most frequent keywords of Flutamide, Nanocomposite, Voltammetry, Anticancer Drug, and Electrophilicity.

The 7$^{th}$ group distinguished by Sarcosine includes four hot keywords of Urine, Electrochemical Biosensor, Oligonucleotides, Label Free Detection, which have seven sensors such as Sarcosine Biosensor, Potentiometric Sensor, Electrochemical Biosensor, Urine Sensors, Graphene-based Electrical Sensor, Disposable Sensor Chips, and Colorimetric Biosensor in their category.

In group 8 related to Apoptosis, we have five keywords of Chemotherapy, DNA Damage, DNA Repair, Rad9, and Tumor Suppressor. Three sensors of Carbon Nanotube Biosensor, Wireless Pressure sensors, and Implantable Pressure Sensor are the important sensing technologies in this group.

### 3.4. Colorectal cancer

Finally, the colorectal cancer co-word network contains 529 nodes and 528 links. Figure 5 shows the co-occurrence network of keywords of this cancer and their connections with sensors.

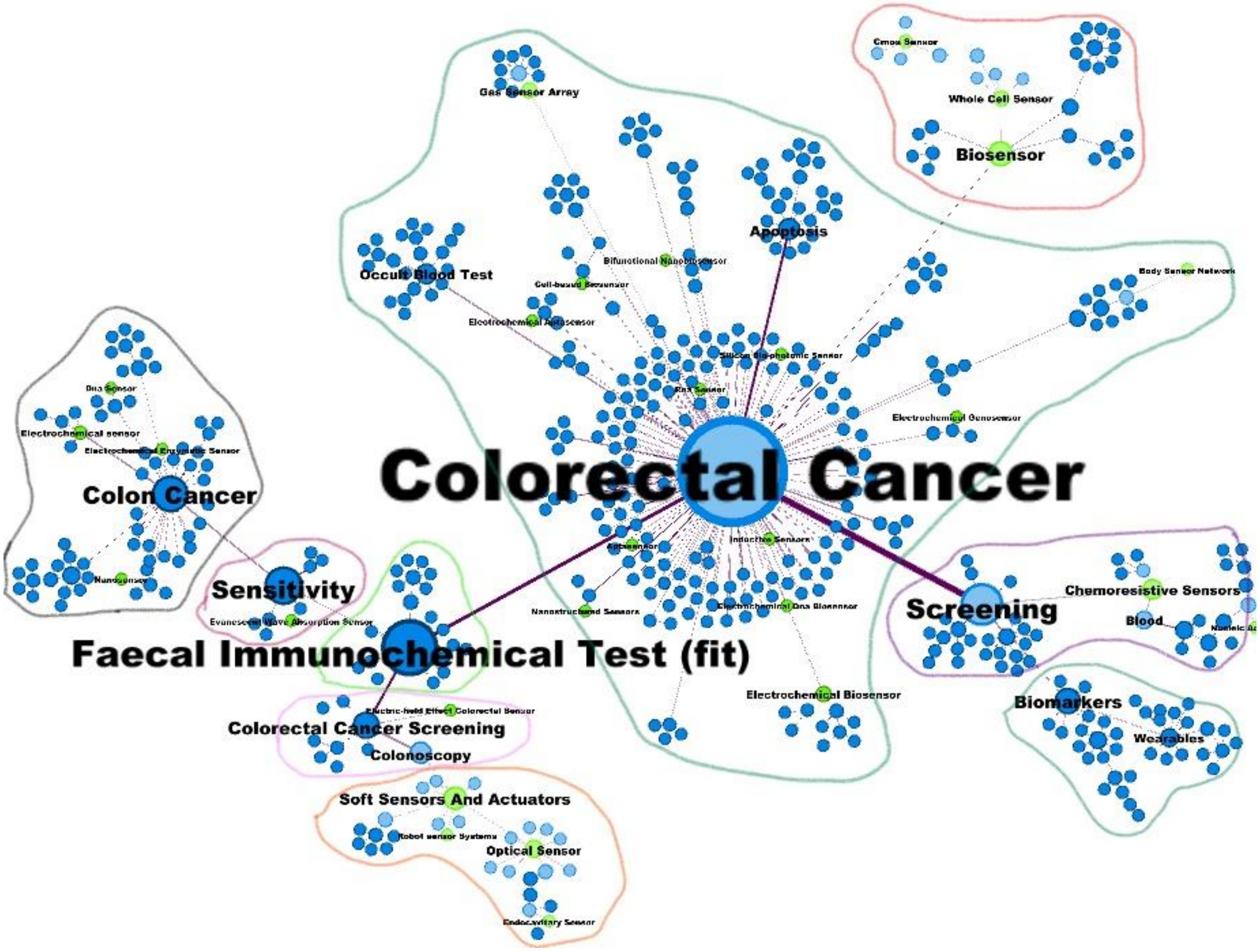

Figure 5. Co-word analysis map of colorectal cancer

Figure 5 shows that there are 9 groups in the colorectal cancer network. The most important nodes based on the Betweenness centrality value are colorectal cancer, Faecal Immunochemical Test, Screening, and Biosensor. Table 4A in Appendix  shows the groups, core keywords of them and related sensors. Table 4A, shows the most important information about groups, core keywords and related sensors in the word co-occurrences network of the colorectal cancer. Also, there are 31 sensors interconnected to this network.

Group 1, led by Colorectal Cancer, is related to five hot topics of Apoptosis, Occult Blood Test, Precision Medicine, Social Health, and Electrochemical Biosensor. It is also illustrated that 13 sensors are interconnected to this path. Electrochemical Biosensor, Gas Sensor Array, Nanostructured Sensors, Bifunctional Nanobiosensor, and Inductive Sensors are a couple of sensors involved in creating linkage path among these nodes.

In group 2, led by Faecal Immunochemical Test (fit), we identified just 5 top keywords of Inflammatory Bowel Disease, Reg3, Quality Assurance, Advance Notification, and Letter.

Group 3 connected to Screening has two interconnected sensors including, Chemoresistive Sensors and Nucleic Acid-sensor. The most frequent keywords involved in this path group are Chemoresistive Sensors, Blood, Tumor Marker, Tissue Microarray, Faecal Occult Blood Test.

Group 4 with the head of Sensitivity is related to important keywords of Antibody, Prostatic Carcinoma, DNA, Acoustic Wave Sensor, Cancer Metastasis. Four sensors of Biosensor, Optical Sensor, Acoustic Wave Sensor, Bio-mems Force Sensor are part of this path group.

The next group, led by Colon Cancer connected to the top five keywords of Zinc, Chromogranin A, Fluorescent Sensor, making Diagnosis, and Peptide relate to four sensors of Trans-rectal Pressure Sensor Array, Chemosensor, and Fluorescent Sensor.

In Group 6, Colorectal Cancer Screening with the most important position in creating the path among different sensors and keywords is responsible for bridging Nanosensor, Flutamide Sensor, Bicalutamide Sensor and most frequent keywords of Flutamide, Nanocomposite, Voltammetry, Anticancer Drug, and Electrophilicity.

The 7$^{th}$ group distinguished by Biomarkers includes five hot keywords of Endoscopy, Au Nps, Signal Enhancement, Erlotinib, and Tollen's Reagent, which have two sensors: Histidine sensor and Fiber Optic Biosensor in their category.

In group 8, related to Biosensor, we have five DNA Mutation, Chemical Sensor, Urine Headspace, Colorimetric Sensor, and Tuberculosis keywords. Ten sensors of Chemical Sensor, Colorimetric Sensor, Cancer Sensor, Array-based Sensor, Temperature Sensors, Impedance-based Sensor, Attachable Gas Sensors, Hg(ii) Sensor, Eis Aptamer Sensor, and Microrna Sensor are the important sensing technologies in this group.

Group 9 led by Soft Sensors and Actuators, is related to five hot topics: Gold Nanoparticle, Immunosensor, Detection, Porcine Liver Esterase, and Fluorescent Probe. It is also illustrated that 27 sensors are interconnected to this path. Immunosensor, Near-infrared Biosensor, Resonance Sensor, Fna-based Electrochemical Sensor, Impedimetric Sensor, and Aptasensor are a couple of sensors involved in the creation of linkage path among these nodes.

According to Table 2, Biosensor is the only type of sensor that plays an essential role in all types of cancer: breast cancer, lung cancer, prostate cancer, and colorectal cancer. After that, the electrochemical sensor is in all types of cancers except lung cancer. Surprisingly, electrochemical biosensor is used in breast cancer, lung cancer, and prostate cancer research but not in colorectal cancer. Optical Sensor can also be considered one of the sensor technologies that significantly is used in three types of cancer:  breast cancer, prostate cancer, and colorectal cancer. This study shows that this type of sensor is applied in more diversified approaches. Moreover, the oxygen sensor ,as a type of gas sensor, is mostly applied in lung cancer and breast cancer studies due to the usage of breath analysis in the treatment process. Cmos Sensor is another technology used in two types of cancer studies, including Lung Cancer and Colorectal Cancer.

Table 2. The most frequent sensors in cancer studies

| Sensor technologies | Breast Cancer | Lung Cancer | Prostate Cancer | Colorectal Cancer |
|---|---|---|---|---|
| Biosensor | * | * | * | * |
| Electrochemical Sensor | * | | * | * |
| Electrochemical Biosensor | * | * | * | |
| Oxygen sensor | * | * | | |
| Tactile Sensor | * | | | |
| Immuno Sensor | * | | | |
| Fiber Optic Sensor | * | | | |
| Cancer Sensor | * | | | |
| Optical Sensor | * | | * | * |
| Pressure Sensors | * | | | |
| Dna Sensor | * | | | |
| Impedimetric Sensor | * | | | |
| Terahertz Sensor | * | | | |
| Genosensor | * | | | |
| Biomimetic Sensor | * | | | |
| Electrochemical Cytosensor | * | | | |
| Gas sensor | | * | | |
| Chemical Sensor | | * | | |
| Fluorescence Sensor | | * | | |
| Cell Sensor | | * | | |
| Metal Oxide Semiconductor Sensor | | * | | |
| Breath Sensor | | * | | |
| Electrochemical Immunosensor | | * | | |
| Capacitive Biosensor | | * | | |
| Exhaled Breath Sensor | | * | | |
| Acetone Sensor | | * | | |
| Volatile Organic Compound Sensors | | * | | |
| Cmos Sensor | | * | | * |
| Electrochemical Genosensor | | * | | |
| Colorimetric Sensor | | * | | |
| Immunosensor | | | * | |
| Near-infrared Biosensor | | | * | |
| Resonance Sensor | | | * | |
| Chemosensor | | | * | |
| Glucose Sensor | | | * | |
| Impedimetric Sensor | | | * | |
| Soft Sensors And Actuators | | | | * |
| Whole Cell Sensor | | | | * |

## 4. Conclusions

The evolution of the ecosystem of sensors technology over the last few decades is unparalleled with an intensive activity of research in public and private laboratories (Andersen et al., 2004; Coccia et al., 2021; Coccia, 2005; 2015, 2017a, 2017b; Pagliaro and Coccia, 2021). Sensor technology is co-evolving with growing interactions of technological systems directed to fulfil human goals and needs and solve problems in society. Cancer is still one of the leading diseases and causes of death in the world. More than 250 types of cancers are currently known (cf., Coccia and Bellitto, 2018). Types of cancer under study here are a major cause of cancer-related deaths globally due to their difficult diagnosis in early stages resulting in late treatment. In fact, in the health domain, a major challenge is the detection of diseases using rapid and cost-effective techniques. Most of the existing cancer detection methods show poor sensitivity and selectivity and are time consuming with high cost (Mohan et al., 2022). In short, the early diagnosis is an important integral part of the process of cancer treatment. For this reason, the analysis of the role of sensor technology and research in these topics is basic in clinical diagnosis and early treatment for patients for reducing the mortality worldwide.

The task of this study was to do an exploratory analysis on the role of sensor technology in cancer research to see possible new directions for improving diagnosis and therapeutic treatments. The results of this analysis are:

1. Biosensor is the only type of sensor that plays an essential role in all types of four cancer under study.
2. Electrochemical sensor is in all types of cancer except lung cancer.
3. Electrochemical Biosensor is used in breast cancer, lung cancer, and prostate cancer research but not colorectal cancer.
4. Optical Sensor is a technology used in three types of cancer: breast cancer, prostate cancer, and colorectal cancer.
5. Oxygen sensor has a role in lung cancer and breast cancer studies due to the usage for breath analysis in the treatment process.
6. Finally, Cmos Sensor is another technology used in two types of cancer studies, including Lung Cancer and Colorectal Cancer.

Overall, then, results suggest that new directions, such as optical biosensors that are rapid, real-time, and portable technology, have a low detection limit and a high sensitivity, and have a great potential for diagnosing various types of cancer. Optical biosensors can detect cancer in a few million malignant cells, in comparison to conventional diagnosis techniques that use 1 billion cells in tumor tissue with a diameter of 7 nm–10 nm (traditional methods that are also costly, inconvenient, complex, time consuming, and require technical specialists;  Kaur et al., 2022). Moreover, the cancer biomarkers using luminescence and electrochemical Metal-organic framework sensors have been opening the way for personalized patient treatments and the development of new cancer-detecting devices (Mohan et al., 2022). The challenge of sensor technology in cancer research is the developing of simple, reliable and sensitive point-of-care testing biosensor for cancerous exosomes detection to early cancer diagnosis and prognosis. The biosensor could also avoid the influence of the external environment, including surrounding light and temperature( Zou et al., 2022).

These conclusions are, of course, tentative. Although this study has provided some interesting, albeit preliminary results, it has several limitations. First, a limitation of this study is that sources understudy may only capture certain aspects of the ongoing dynamics of sensor research and technology in cancer research. Second, there are multiple confounding factors that could have an important role in the interaction between sensor technology  and cancer research for diagnosis to be further investigated, such as high R&D investments, collaboration intensity, openness, intellectual property rights, etc. (Roshani et al., 2021). Third, the computational and statistical analyses in this study focus on data in a specific period and should be extended to other periods. Forth, sensor research associated with cancer studies change their borders during the evolution of science, such that the identification of stable technological trajectories and new patterns in the evolution of sensors  in cancer research is a non-trivial exercise.

To conclude, future research should consider new data when available, and when possible, apply new approaches to reinforce proposed results. Despite these limitations, the results presented here clearly illustrate the evolutionary paths of main sensor technologies that have a great potential as a powerful tool in future diagnosis of cancer but we also need a detailed examination of other aspects and factors for supporting appropriate strategies of research and innovation policy, and management of technology to foster the technology transfer of sensor in cancer research for improving diagnosis and at the same time  reducing, as far as possible, world-wide mortality of cancer ins society.

## APPENDIX

**Table 1A. Groups, Core keywords and related sensors in the breast cancer network**

| Group | Core Keyword | Top 5 keywords | Related sensors |
|---|---|---|---|
| **1** | Breast Cancer | Autophagy<br>Immunoassay<br>Electrochemical Biosensor<br>Tumorigenesis<br>Microrna | 1. Electrochemical Biosensor<br>2. Electrochemical Sensor<br>3. Oxygen sensor<br>4. Immuno Sensor<br>5. Array-based Sensor<br>6. Terahertz Sensor<br>7. Dna Sensor<br>8. Fiber Optic Sensor<br>9. Impedimetric Sensor<br>10. Fluorescence Sensor<br>11. Dna Biosensor<br>12. Single-use Sensors<br>13. Impedance Sensor<br>14. Photoelectrochemical Biosensor<br>15. Electrochemical Cytosensor<br>16. Optical Biosensor<br>17. Antifouling Biosensor<br>18. Electrochemical Dna Sensor<br>19. Electrochemical Dna Biosensor<br>20. Cytosensor<br>21. Surface Plasmon Resonance Biosensor<br>22. Micro Sensor<br>23. Dna Damage Sensor<br>24. Nuclease Optical Sensor<br>25. Microfluidic Immunosensor<br>26. Gan Hemt Based Biosensor<br>27. Aptasensor<br>28. Chemical Sensor<br>29. Radio Sensor Technology (rst)<br>30. Miniature Sensor<br>31. Electrochemical Genosensor<br>32. Cell-based Biosensor<br>33. Metabolic Sensor<br>34. Multimodal Sensors<br>35. Multi-modal sensor Data<br>36. Acoustic Biosensors<br>37. Silicon-sensor Chips<br>38. Electrochemical Aptasensor<br>39. Estrogen Biosensor<br>40. Stress Sensor<br>41. Spr-based Pcf Sensor<br>42. Oil Adulteration sensor<br>43. New Ion-channel Sensor Model<br>44. Environmental Sensor<br>45. Fluorescent Biosensor<br>46. Body Sensor Network<br>47. Breath Sensor |

48. Optical 3-d Sensor
49. Mems Sensor
50. Nanobiosensor
51. Quartz Crystal Microbalance Sensor
52. Aunps Sensor
53. Nanomechanosensor
54. Ratiometric Aptasensor
55. Recyclable Sensor
56. Redox Sensor
57. Thermal Sensors
58. Bionic Sensor
59. Mems Mass Sensor
60. Genomagnetic Sensor
61. on-off Type Aptasensor
62. Piezoresistive Force Sensor
63. Sandwich-type Immunosensor
64. Dsrna sensor
65. Cmos Capacitance Sensor
66. Plasmonic Biosensors
67. Cell Sensor
68. E-DNA Sensor
69. Metal Nanoparticles-graphene Hybrid Biosensors
70. Multiplex Dna Sensor
71. Genetically-encoded Sensor
72. Apoptosis Sensor
73. Lectin Sensor
74. Glucose Biosensor
75. Molecular Sensors
76. Electrochemical Sandwich Immunosensor
77. sensor Activity
78. Label-free Electrochemical Sensor
79. Self-powered Sensor
80. Bio-mems Force Sensor
81. Sandwich-type Biosensor
82. Fiber Optic Biosensor
83. Fluorescent Sensor
84. Aptamersensor
85. Gold Nanostructured Sensor
86. Luminescent Sensor
87. Electronic Sensor
88. Laser Distance Sensor
89. Active Radio Sensor
90. Stimulus Response sensor
91. Lspr sensor Multiplexed Detection
92. Plow-through Biosensor
93. Fluorescence Turn-on Sensor
94. Rhodamine-coumarin Based Chemo sensor
95. Biochemosensor
96. Sh-saw Biosensor
97. Electrochemiluminescence Immunosensor
98. Live Cell Non-invasive Apoptosis Detection Sensor (niads)
99. Breast Cancer Sensor

| | | | |
|---|---|---|---|
| | | | 100.Kinase Biosensor |
| **2** | Biosensor | Her2<br>Cancer antigen<br>Nanoparticle<br>Tactile Sensor<br>Signal Amplification | 1- Biosensor<br>2- Optical Sensor<br>3- Colorimetric Sensor<br>4- Colorimetric Nanosensor<br>5- Ph. Sensor<br>6- Refractive Index Sensor<br>7- Nanosensor<br>8- Label-free Optical Sensor<br>9- Label-free Aptasensor<br>10- Electrogenerated Chemiluminescence Aptasensor<br>11- Fluid-type Tactile Sensor<br>12- Vision-based Sensor<br>13- Spr Sensor<br>14- Label-free Electrochemical Immunosensor<br>15- Silicon Nanobiosensor<br>16- Protease Sensor<br>17- Environmental Sensor<br>18- Biochemosensor<br>19- Tactile Sensor<br>20- Pressure Sensors<br>21- Genosensor<br>22- Cancer Sensor<br>23- Biomimetic Sensor<br>24- Mirna sensor<br>25- Electrochemical Immunosensor<br>26- Occipital Structure Sensor<br>27- Multiplexed Immunosensor<br>28- Acoustic Sensor<br>29- Ratiometric Electrochemical Biosensor<br>30- Lspr Biosensor<br>31- Atomic Force Microscopy Sensor |
| **3** | Reactive Oxygen Species | Apoptosis<br>Mcf-7 Cancer Cells<br>Circulating Tumor Cell<br>Dna Hydrogel Biosensor<br>Carbon Dot | 1) Raman Biosensor<br>2) Dna Hydrogel Biosensor<br>3) Light Addressable Potentiometric Sensor (laps)<br>4) Capacitive Sensor<br>5) Label-free Biosensor<br>6) Nanowire Biosensor<br>7) Motion Sensor<br>8) Targeted Drug Delivery sensor<br>9) Ratiometric Fluorescent Sensor<br>10) Hall Sensor<br>11) Cell-based Sensor<br>12) Metal Ion Sensors<br>13) Conductivity Sensor<br>14) Turn Off Fluorescence Sensor<br>15) Hydrogel Sensor<br>16) Magnetoelastic Sensor<br>17) Cmos Image Sensor<br>18) Pb2+ Ions sensor |

**Table 2A. Groups, Core keywords and related sensors in the lung cancer network**

| Group | Core Keyword | Top 5 keywords | Related sensors |
|---|---|---|---|
| **1** | Lung Cancer | Electronic Nose (e-nose)<br>Pattern Recognition<br>Mutation<br>Nrf2 (Nuclear factor-erythroid factor 2-related factor 2)<br>Calix[4]arene | 1. Oxygen sensor<br>2. Electrochemical Biosensor<br>3. Cell Sensor<br>4. Metal Oxide Semiconductor Sensor<br>5. Chemiresisitve Sensor<br>6. Breath Sensor<br>7. Electrochemical Immunosensor<br>8. Capacitive Biosensor<br>9. Photoelectrochemical Immunosensor<br>10. Vapor sensor<br>11. Ammonia Gas Sensor<br>12. Voc Gas Sensor<br>13. Electrochemical Sensor<br>14. Mems Magnetic Sensor<br>15. Immunosensor<br>16. Sensor-based Gas Analyzer<br>17. Nonconjugated Polymer Gas Sensor<br>18. Color Sensor<br>19. Computed Tomography-guided Sensor Implantation<br>20. Metal Oxide Mox Sensor<br>21. Dna Damage Sensor Protein<br>22. Sensor Phenomena And Characterization<br>23. Label-free Electrochemical Immunosensor<br>24. Mox Sensors<br>25. Low sensor Chamber Volume<br>26. Metal Oxide Sensors<br>27. Artificial Olfactory Sensor<br>28. Optical Fiber Sensors<br>29. Optical Sensor<br>30. Electronic Smell Sensor<br>31. Colorimetric Cross-responsive Sensor<br>32. Sensor-type Prototype System<br>33. Mos Sensors<br>34. Sensor Selection<br>35. Fluorescence Turn-on Chemosensor<br>36. Coumarin Based Sequential Chemosensor<br>37. Nanostructured sensor Materials<br>38. Love-wave Sensor<br>39. Calorimetric Sensor<br>40. Gas-sensor Property<br>41. Electrokinetic Sensor<br>42. Chemosensors<br>43. Surface Stress Sensor<br>44. Nanomechanical Sensor<br>45. Label-free Immunosensor<br>46. Chemical Gas Sensor<br>47. Mach Zehnder Interferometer Sensor<br>48. Picric Acid Sensor<br>49. Diverse Sensor Array<br>50. Virtual Sensors Array |
| | | | 51. Aptasensor<br>52. Label-free Nanoimmunosensor<br>53. Microarray Sers Sensors<br>54. Capacitive Sensor<br>55. Nh3 Gas Sensor<br>56. Smartphone-integrated Sensor<br>57. sensor For Health<br>58. Silicon Biophotonic Sensor<br>59. Nanobiosensor<br>60. Single Molecule sensor<br>61. Fluoride Ions sensor<br>62. Breath Ethanol Sensors<br>63. Silicon Nanobiosensor<br>64. Poly-silicon Wire Sensor<br>65. Cytosensor<br>66. Adsorption-based Sensor<br>67. Graphene Biosensor<br>68. Alkane Sensor<br>69. Luminescent Sensor |
| **2** | Volatile Organic Compounds (voc) | Breath<br>Diagnosis<br>Gold Nanoparticle<br>Nanoparticles<br>Exhaled Breath | 1- Exhaled Breath Sensor<br>2- Dna Sensor<br>3- Volatile Organic Compound Sensors<br>4- Cmos Sensor<br>5- Electrochemical Genosensor<br>6- Metal Oxide Semiconductor Gas Sensor<br>7- Intracellular Sensor<br>8- Amperometric Immunosensor<br>9- Odor Sensors<br>10- Chemical Piezosensor<br>11- Chip Integrated Biosensor<br>12- Nanophotonic Biosensor<br>13- Stochastic Sensor<br>14- Lectin Sensor<br>15- Hydrogen Gas Sensor<br>16- Impedance Biosensor<br>17- Alcohol Nanosensor<br>18- Paper-based Sensors<br>19- Chemoresistive Sensors<br>20- Multianalyte Biosensor<br>21- Co2 Biosensor<br>22- Characterization Of The Sensor<br>23- Trace Gas sensor |
| **3** | Breath analysis | Real Time<br>Diabetes<br>Non-invasive<br>Formaldehyde<br>Sno2 | 1- Piezoresistive Membrane Sensors<br>2- H2s Sensor<br>3- Formaldehyde Gas Sensor<br>4- Quantum Resistive Vapor Sensor<br>5- Quantum Resistive Sensor<br>6- Acetone Sensor<br>7- Genetically Encoded Nanosensor<br>8- Gas Biosensor |
| **4** | Gas Sensor | Humidity sensor<br>Health Monitoring | 1- Gas sensor<br>2- Lung Cancer Sensor |
| | | Reduced Graphene Oxide<br>Acetone<br>Tin Oxide | 3- Multiparameter Virtual Sensor Array<br>4- Humidity sensor<br>5- sensor Mechanism<br>6- Flexible Sensor |
| **5** | Biosensor | Wireless Sensor Networks<br>Radon<br>Smart Home<br>Electrochemical<br>Inhibitors | 1- Biosensor<br>2- Fluorescence Biosensor<br>3- Wireless Sensor Networks<br>4- Photoelectric Aerosol Sensor<br>5- Direct Progeny Sensor<br>6- Pid Sensor<br>7- Chemo-resistive Gas Sensor<br>8- Hydrocarbon Gas Sensor |
| **6** | Exosome | Cancer Diagnosis<br>Optical Biosensor<br>Immunoassay<br>Multi-wall Carbon<br>Nanotubes | 1- Optical Biosensor<br>2- Saw Gas Sensor<br>3- Optical Chemical Sensor, Vocs Identification<br>4- Saw Immunosensor<br>5- Impedimetric Sensor |
| **7** | Surface Plasmon Resonance (spr) | Endoscopy<br>Au Nps<br>Signal Enhancement<br>Erlotinib<br>Tollen's Reagent | 1- Histidine sensor<br>2- Fiber Optic Biosensor |
| **8** | Graphene | Dna Mutation<br>Chemical Sensor<br>Urine Headspace<br>Colorimetric Sensor<br>Tuberculosis | 1- Chemical Sensor<br>2- Colorimetric Sensor<br>3- Cancer Sensor<br>4- Array-based Sensor<br>5- Temperature Sensors<br>6- Impedance-based Sensor<br>7- Attachable Gas Sensors<br>8- Hg(ii) Sensor<br>9- Eis Aptamer Sensor<br>10- Microrna Sensor |

**Table 3A. Groups, Core keywords and related sensors in the Prostate cancer network**

| Group | Core Keyword | Top 5 keywords | Related sensors |
|---|---|---|---|
| **1** | Prostate cancer | Gold Nanoparticle<br>Immunosensor<br>Detection<br>Porcine Liver Esterase<br>Fluorescent Probe | 5. Immunosensor<br>6. Near-infrared Biosensor<br>7. Resonance Sensor<br>8. Fna-based Electrochemical Sensor<br>9. Impedimetric Sensor<br>10. Aptasensor<br>11. Piezoelectric Sensor<br>12. Boronate Sensor<br>13. Zinc sensor<br>14. Nanopore-based Sensor<br>15. Nanowire Biosensor<br>16. Microsensors<br>17. Voltammetric Sensor<br>18. Hall Sensor<br>19. Electrochemical Aptasensor<br>20. Tactile Sensor<br>21. Array-based Sensor<br>22. Non-enzymatic Sensor<br>23. Electrochemical Aptamer Sensors<br>24. Ratiometric Aptasensor<br>25. Light Addressable Potentiometric Sensor (laps)<br>26. Camera Sensor<br>27. Fluorescence Gas-sensory Arrays |
| **2** | Activated Protein Kinase (Ampk) | Apoptosis<br>Lkb1<br>Lung Cancer<br>Mtorc1<br>Castrate-resistant | 1. Label-free Nanoimmunosensor |
| **3** | Prostate Specific Antigen (psa) | Aptamer<br>Prostate Cancer Diagnosis<br>Cancer Biomarker<br>Silicon Nanowire | 1. Glucose Sensor<br>2. Ph Sensor<br>3. Colorimetric Sensor<br>4. Quantum Resistive Sensor<br>5. Label-free Biosensors<br>6. Love Wave Sensor<br>7. Biomarker sensor<br>8. Photonic Sensor<br>9. Biomarker Pressure Sensor<br>10. Carbon Nanotube Sensors<br>11. Ion Selective Sensors<br>12. Rna Biosensor<br>13. Affinity Sensor<br>14. Electrical Conductivity Sensor<br>15. Conductivity Sensor<br>16. Memristive Biosensors<br>17. Ring Resonator Biosensor<br>18. Nanobiosensor<br>19. Metal Nanoparticles-graphene Hybrid Biosensors<br>20. Impedimetric Immunosensor<br>21. Single-molecule Sensor |
| | | | 22. Sandwich-type Electrochemical Immunosensor |
| **4** | Biosensor | Antibody<br>Prostatic Carcinoma<br>Dna<br>Acoustic Wave Sensor<br>Cancer Metastasis | 1. Biosensor<br>2. Optical Sensor<br>3. Acoustic Wave Sensor<br>4. Bio-mems Force Sensor |
| **5** | Prostate | Zinc<br>Chromogranin A<br>Fluorescent Sensor<br>Imaging Diagnosis<br>Peptide | 1. Trans-rectal Sensor<br>2. Pressure Sensor Array<br>3. Chemosensor<br>4. Fluorescent Sensor |
| **6** | Electrochemical Sensor | Flutamide<br>Nanocomposite<br>Voltammetry<br>Anticancer Drug<br>Electrophilicity | 1. Electrochemical Sensor<br>2. Nanosensor<br>3. Flutamide Sensor<br>4. Bicalutamide Sensor |
| **7** | Sarcosine | Urine<br>Electrochemical Biosensor<br>Oligonucleotides<br>Label Free Detection | 1. Sarcosine Biosensor<br>2. Potentiometric Sensor<br>3. Electrochemical Biosensor<br>4. Urine Sensors<br>5. Graphene-based Electrical Sensor<br>6. Disposable Sensor Chips<br>7. Colorimetric Biosensor |
| **8** | Apoptosis | Chemotherapy<br>Dna Damage<br>Dna Repair<br>Rad9<br>Tumor Suppressor | 1. Carbon Nanotube Biosensor<br>2. Wireless Pressure Sensor<br>3. Implantable Pressure Sensor |

**Table 4A. Groups, Core keywords and related sensors in the colorectal cancer network**

| Group | Core Keyword | Top 5 keywords | Related sensors |
|---|---|---|---|
| **1** | Colorectal Cancer | Apoptosis<br>Occult Blood Test<br>Precision Medicine<br>Social Health<br>Electrochemical Biosensor | 4. Electrochemical Biosensor<br>5. Gas Sensor Array<br>6. Nanostructured Sensors<br>7. Bifunctional Nanobiosensor<br>8. Inductive Sensors<br>9. Body Sensor Network<br>10. Electrochemical Genosensor<br>11. Electrochemical Dna Biosensor<br>12. Silicon Bio-photonic Sensor<br>13. Aptasensor<br>14. Cell-based Biosensor<br>15. Electrochemical Aptasensor<br>16. Rna Sensor |
| **2** | Faecal Immunochemical Test (fit) | Inflammatory Bowel Disease<br>Reg3<br>Quality Assurance<br>Advance Notification Letter | - |
| **3** | Screening | Chemoresistive Sensors<br>Blood<br>Tumor Marker<br>Tissue Microarray<br>Faecal Occult Blood Test | 1. Chemoresistive Sensors<br>2. Nucleic Acid-sensor |
| **4** | Sensitivity | Stability<br>Specificity | 1. Evanescent Wave Absorption Sensor |
| **5** | Colon Cancer | Transmission Electron Microscopy (tem)<br>Oxidative Stress<br>Kras<br>Drug Screening | 1- Electrochemical Sensor<br>2- Electrochemical Enzymatic Sensor<br>3- Dna Sensor<br>4- Nanosensor |
| **6** | Colorectal Cancer Screening | Colonoscopy<br>Quantitative Fecal Immunochemical Test For Hemoglobin<br>Sample Stability | 1- Electric-field Effect Colorectal Sensor |
| **7** | Biomarkers | Wearables<br>Sex<br>Circadian Rhythms<br>Nanotechnology<br>Hemolysis Assay | - |
| **8** | Biosensor | Cancer Markers<br>Surface Plasmon Resonance<br>Whole Cell Sensor<br>Circular Dichroism Spectroscopy | 1. Biosensor<br>2. Whole Cell Sensor<br>3. Cmos Sensor |
| **9** | Soft Sensors And Actuators | Optical Sensor<br>Bowel Viability<br>Colon<br>Cloud Serve<br>Pulse Oximetry | 1. Optical Sensor<br>2. Robot sensor Systems<br>3. Endocavitary Sensor |

**References**

Andersen P. D., Jørgensen B. H., Lading L., Rasmussen B. 2004. Sensor foresight-technology and market. Technovation, vol. 24, n. 4, pp. 311-320, https://doi.org/10.1016/S0166-4972(02)00072-X.

Ardito L., Coccia M., Messeni Petruzzelli A. 2021. Technological exaptation and crisis management: Evidence from COVID-19 outbreaks. R&D Management, vol. 51, n. 4, pp. 381-392. https://doi.org/10.1111/radm.12455

Bastian, M., Heymann, S., & Jacomy, M. 2009. Gephi: an open source software for exploring and manipulating networks. In Third international AAAI conference on weblogs and social media.

Bax, C., Prudenza, S., Gaspari, G., (...), Grizzi, F., Taverna, G.2022. Drift compensation on electronic nose data for non-invasive diagnosis of prostate cancer by urine analysis, iScience25(1),103622

Bayford, R.H., Damaso, R., Neshatvar, N., (...), Nordebo, S., Demosthenous, A. 2022. Locating Functionalized Gold Nanoparticles Using Electrical Impedance Tomography, IEEE Transactions on Biomedical Engineering 69(1), pp. 494-502

Chagpar A. B., Coccia M. 2019. Factors associated with breast cancer mortality-per-incident case in low-to-middle income countries (LMICs). Journal of Clinical Oncology, vol. 37, no. 15, suppl. pp. 1566-1566, DOI: 10.1200/JCO.2019.37.15

Coccia M. 2005. A taxonomy of public research bodies: a systemic approach, Prometheus, vol. 23, n. 1, pp. 63-82. https://doi.org/10.1080/0810902042000331322

Coccia M. 2012. Evolutionary growth of knowledge in path-breaking targeted therapies for lung cancer: radical innovations and structure of the new technological paradigm. International Journal of Behavioural and Healthcare Research, vol. 3, nos. 3-4, pp. 273-290. https://doi.org/10.1504/IJBHR.2012.051406

Coccia M. 2013. The effect of country wealth on incidence of breast cancer, Breast Cancer Research and Treatment, vol. 141, n. 2, pp. 225-229, https://doi.org/10.1007/s10549-013-2683-y

Coccia M. 2014. Path-breaking target therapies for lung cancer and a far-sighted health policy to support clinical and cost effectiveness, Health Policy and Technology, vol. 1, n. 3, pp. 74-82. https://doi.org/10.1016/j.hlpt.2013.09.007

Coccia M. 2015. Spatial relation between geo-climate zones and technological outputs to explain the evolution of technology. Int. J. Transitions and Innovation Systems, vol. 4, nos. 1-2, pp. 5-21, http://dx.doi.org/10.1504/IJTIS.2015.074642

Coccia M. 2017. Sources of technological innovation: Radical and incremental innovation problem-driven to support competitive advantage of firms. Technology Analysis & Strategic Management, vol. 29, n. 9, pp. 1048-1061, https://doi.org/10.1080/09537325.2016.1268682

Coccia M. 2017a. Disruptive firms and industrial change, Journal of Economic and Social Thought, vol. 4, n. 4, pp. 437-450, http://dx.doi.org/10.1453/jest.v4i4.1511

Coccia M. 2017b. Varieties of capitalism's theory of innovation and a conceptual integration with leadership-oriented executives: the relation between typologies of executive, technological and socioeconomic performances. Int. J. Public Sector Performance Management, Vol. 3, No. 2, pp. 148–168. https://doi.org/10.1504/IJPSPM.2017.084672

Coccia M. 2018. General properties of the evolution of research fields: a scientometric study of human microbiome, evolutionary robotics and astrobiology, Scientometrics, vol. 117, n. 2, pp. 1265-1283, https://doi.org/10.1007/s11192-018-2902-8

Coccia M. 2018a. An introduction to the methods of inquiry in social sciences, J. Adm. Soc. Sci., vol. 5, n. 2, pp. 116-126, http://dx.doi.org/10.1453/jsas.v5i2.1651

Coccia M. 2019. A Theory of classification and evolution of technologies within a Generalized Darwinism, Technology Analysis & Strategic Management, vol. 31, n. 5, pp. 517-531, http://dx.doi.org/10.1080/09537325.2018.1523385

Coccia M. 2019a. The theory of technological parasitism for the measurement of the evolution of technology and technological forecasting, Technological Forecasting and Social Change, vol. 141, pp. 289-304, https://doi.org/10.1016/j.techfore.2018.12.012

Coccia M. 2019b. Why do nations produce science advances and new technology? Technology in society, vol. 59, November, n. 101124, pp. 1-9, https://doi.org/10.1016/j.techsoc.2019.03.007

Coccia M. 2019c. Socioeconomic Analysis of Breast Cancer between Countries, 2012-2018 Period. Quaderni IRCrES, 4(2), 37-46. http://dx.doi.org/10.23760/2499-6661.2019.009

Coccia M. 2019d. Artificial intelligence technology in cancer imaging: Clinical challenges for detection of lung and breast cancer. Journal of Social and Administrative Sciences, vol. 6, n. 2, pp. 82-98, http://dx.doi.org/10.1453/jsas.v6i2.1888

Coccia M. 2020. Deep learning technology for improving cancer care in society: New directions in cancer imaging driven by artificial intelligence. Technology in Society, vol. 60, February, pp. 1-11, art. n. 101198, https://doi.org/10.1016/j.techsoc.2019.101198

Coccia M. 2021. Technological Innovation. The Blackwell Encyclopedia of Sociology. Edited by George Ritzer and Chris Rojek. John Wiley & Sons, Ltd .DOI: 10.1002/9781405165518.wbeost011.pub2

Coccia M. 2021a. Evolution of technology in replacement of heart valves: Transcatheter aortic valves, a revolution for management of valvular heart diseases, Health Policy and Technology, vol. 10, Issue 2, June 2021, Article number 100512, https://doi.org/10.1016/j.hlpt.2021.100512

Coccia M., Bellitto M. 2018. Human progress and its socioeconomic effects in society, Journal of Economic and Social Thought, vol. 5, n. 2, pp. 160-178, ISSN: 2149-0422, www.kspjournals.org, http://dx.doi.org/10.1453/jest.v5i2.1649

Coccia M., Finardi U. 2012. Emerging nanotechnological research for future pathways of biomedicine. International Journal of Biomedical nanoscience and nanotechnology, vol. 2, nos. 3-4, pp. 299-317. DOI: 10.1504/IJBNN.2012.051223

Coccia M., Finardi U. 2013. New technological trajectories of non-thermal plasma technology in medicine. Int. J. Biomedical Engineering and Technology, vol. 11, n. 4, pp. 337-356, DOI: 10.1504/IJBET.2013.055665

Coccia M., Roshani S., Mosleh M. 2021. Scientific Developments and New Technological Trajectories in Sensor Research. Sensors, vol. 21, no. 23: art. N. 7803. https://doi.org/10.3390/s21237803

Coccia M., Watts J. 2020. A theory of the evolution of technology: technological parasitism and the implications for innovation management, Journal of Engineering and Technology Management, vol. 55 (2020) 101552, https://doi.org/10.1016/j.jengtecman.2019.11.003

Delecroix, B. and Epstein, R., 2004. Co-word analysis for the non-scientific information example of Reuters Business Briefings. Data Science Journal, 3, pp.80-87.

Globocan 2020. Cancer today. Estimated age-standardized incidence and mortality rates (World) in 2020, worldwide, both sexes, all ages. https://gco.iarc.fr/ (accessed 3 February 2022)

Harbeck, N., Gnant, M., Breast cancer, (2017) The Lancet, 389 (10074), pp. 1134-1150

Jiang, X., Wang, C., Ke, Z., (...), Yang, Y., Dai, Y. 2022. The ion channel TRPV1 gain-of-function reprograms the immune microenvironment to facilitate colorectal tumorigenesis. Cancer Letters, 527, pp. 95-106

Joshi, S., Guruprasad, G., Kulkarni, S., Ghosh, R. 2022. Reduced Graphene Oxide Based Electronic Sensors for Rapid and Label-Free Detection of CEA and CYFRA 21-1, IEEE Sensors Journal 22(2), pp. 1138-1145

Kashani, E. S., & Roshani, S. 2019. Evolution of innovation system literature: Intellectual bases and emerging trends. Technological Forecasting and Social Change, 146, 68-80.

Kaur, B., Kumar, S., Kaushik, B.K. 2022. Recent advancements in optical biosensors for cancer detection. Biosensors and Bioelectronics, 197,113805

Kaya, S.I., Ozcelikay, G., Mollarasouli, F., Bakirhan, N.K., Ozkan, S.A. 2022. Recent achievements and challenges on nanomaterial based electrochemical biosensors for the detection of colon and lung cancer biomarkers. Sensors and Actuators B: Chemical 351,130856

Kim, G., Wu, Q., Chu, J.L., (...), Moore, J.S., Li, K.C. 2022. Ultrasound controlled mechanophore activation in hydrogels for cancer therapy, Proceedings of the National Academy of Sciences of the United States of America 119(4),e2109791119

Li, B. , Pan, W. , Liu, C. 2020. Homogenous Magneto-Fluorescent Nanosensor for Tumor-Derived Exosome Isolation and Analysis ) ACS Sensors

Li, B., Liu, C. , Pan, W. 2020a. Facile fluorescent aptasensor using aggregation-induced emission luminogens for exosomal proteins profiling towards liquid biopsy. Biosensors and Bioelectronics

Lu, J.-Y., Chen, Q.-Y., Meng, S.-C., Feng, C.-J. 2022. A dye-andrographolide assembly as a turn-on sensor for detection of phthalate in both cells and fishAnalytica Chimica Acta 1195,339460

Mohan, B., Kumar, S., Xi, H., (...), Zhang, Y., Ren, P. 2022. Fabricated Metal-Organic Frameworks (MOFs) as luminescent and electrochemical biosensors for cancer biomarkers detection. Biosensors and Bioelectronics, 197,113738

Pagliaro M., Coccia M. 2021. How self-determination of scholars outclasses shrinking public research lab budgets, supporting scientific production: a case study and R&D management implications. Heliyon. vol. 7, n. 1, e05998. https://doi.org/10.1016/j.heliyon.2021.e05998

Pothipor, C., Bamrungsap, S., Jakmunee, J., Ounnunkad, K. 2022. A gold nanoparticle-dye/poly(3-aminobenzylamine)/two dimensional MoSe2/graphene oxide electrode towards label-free electrochemical biosensor for simultaneous dual-mode detection of cancer antigen 15-3 and microRNA-21. Colloids and Surfaces B: Biointerfaces, 210,112260

Prema, P., Boobalan, T., Arun, A., (...), Nguyen, V.-H., Balaji, P. 2022. Green tea extract mediated biogenic synthesis of gold nanoparticles with potent anti-proliferative effect against PC-3 human prostate cancer cells, Materials Letters 306,130882

Quirin, A., Cordón, O., Guerrero-Bote, V. P., Vargas-Quesada, B., & Moya-Anegón, F. 2008. A quick MST-based algorithm to obtain Pathfinder networks (∞, n− 1). Journal of the American Society for Information Science and Technology, 59(12), 1912-1924.

Rao N. S.V., Brooks R.R., Wu C.Q. 2018. Proceedings of International Symposium on Sensor Networks, Systems and Security -Advances in Computing and Networking with Applications, Springer.

Rey-Barth, S., Pinsault, N., Terrisse, H., (...), Guyot, C., Bosson, J.-L. 2022. A program centered on smart electrically assisted bicycle outings for rehabilitation after breast cancer: A pilot studyMedical Engineering and Physics, 100,103758

Roshani S., Bagheri R., Mosleh M., Coccia M. 2021. What is the relationship between research funding and citation-based performance? A comparative analysis between critical research fields. Scientometrics. https://doi.org/10.1007/s11192-021-04077-9

Sci2 Team 2009. Science of science (Sci2) tool. Indiana University and SciTech Strategies, 379.

Sivanandhan, M., Parasuraman, A., Surya, C., (...), Mani, D., Ahn, Y.-H. 2022. Facile approach for green synthesis of fluorescent carbon dots from Manihot esculenta and their potential applications as sensor and bio-imaging agents. Inorganic Chemistry Communications 137,109219

Thakare, S., Shaikh, A., Bodas, D., Gajbhiye, V.2022. Application of dendrimer-based nanosensors in immunodiagnosis. Colloids and Surfaces B: Biointerfaces 209,112174

Tumuluru, P., Hrushikesava Raju, S., Santhi, M.V.B.T., (...), Seetha Rama Krishna, P., Koujalagi, A.2022. Smart Lung Cancer Detector Using a Novel Hybrid for Early Detection of Lung Cancer, Lecture Notes in Networks and Systems, 311, pp. 849-862

Web of Science, WOS 2021. Documents. Clarivate.

Welz, L., Kakavand, N., Hang, X., (...), Rosenstiel, P., Aden, K. 2022. Epithelial X-Box Binding Protein 1 Coordinates Tumor Protein p53-Driven DNA Damage Responses and Suppression of Intestinal Carcinogenesis. Gastroenterology, 162(1), pp. 223-237.e11.

Wilson, J.S., 2004. Sensor technology handbook. Elsevier.

Wu, Y., Feng, Y., Li, X.2022. Classification of breast cancer by a gold nanoparticle based multicolor fluorescent aptasensor, Journal of Colloid and Interface Science 611, pp. 287-293

Zou, L., Liu, X., Zhou, Y., (...), Yang, X., Wang, K. 2022. Optical fiber amplifier and thermometer assisted point-of-care biosensor for detection of cancerous exosomes. Sensors and Actuators B: Chemical, 351,130893